\documentclass[%
 reprint,
nofootinbib,
 amsmath,amssymb,
 aps,
]{revtex4-1}

\usepackage{graphicx}
\usepackage{dcolumn}
\usepackage{bm}
\usepackage{hyperref}

\usepackage{csquotes} 

\usepackage{xcolor}

\renewcommand{\eprint}[1]{\href{http://arxiv.org/abs/#1}{{\tt [arXiv:#1]}}}

\providecommand\jcap{JCAP}

\providecommand\prd{Physical Review D}

\providecommand\apjl{Astrophys.J.Lett.}

\providecommand\mnras{M.N.R.A.S.}

\newcommand{\ie}{{\em i.e.} \/}
\newcommand{\eg}{{\em e.g.} \/}

\newcommand{\cf}{{\em cf.} \/}

\begin{document}

\title{Imprints of decaying dark matter on cosmic voids}

\author{Earl Lester}
\email{earl.sullivanlester@utas.edu.au}
\author{Krzysztof Bolejko}
\email{krzysztof.bolejko@utas.edu.au}
\affiliation{School of Natural Sciences, College of Sciences and Engineering, University of Tasmania, Private Bag 37, Hobart TAS 7001, Australia}


\begin{abstract}
The Standard Cosmological Model assumes that more than 85\% of matter is in the form of collisionless and pressureless dark matter. Unstable decaying dark matter has been proposed in the literature as an extension to the standard cold dark matter model. In this paper we investigate a scenario when dark matter decays and the resultant particle moves with respect to the dark matter.  
A covariant hydrodynamical model is developed in which the decay is modeled by the transfer of energy-momentum between two dark dust fluid components. We parameterise the model in terms of the decay rate $\Gamma$ and injection velocity $v_i$ of the resultant dark matter particles. We apply the framework to study the evolution of cosmic voids which are environments with low content of baryonic matter. Thus, unlike baryon-rich environments, voids provide an opportunity to measure dark matter signals that are less contaminated by complex baryonic processes.  
We find that the growth of S-type voids is modified by the dark matter decay, leading to imprints at the present day.
This paper serves as a proof-of-concept that cosmic voids can be used to study dark mater physics. We argue that future cosmological observations of voids should focus on signs of reported features to either confirm or rule out the decaying dark matter scenario. Lack of presence of reported features could put constraints of the decay of dark matter in terms of $\Gamma > H_0^{-1}$ and  $v_i<10$ km/s.
\end{abstract}

\maketitle

\section{Introduction}

The status of Dark Matter (DM) is as an open problem in modern physics. As DM is known to form the majority of the universal matter content, its presence plays a important part in cosmological modelling and observations. Yet, constraints on the possible particle properties of DM via direct or indirect detection experiments have been inconclusive \cite{mack2014known}, and DM particles have never been directly detected in an experiment on Earth \cite{2013arXiv1309.0528C,2014PhRvL.112i1303A}. Cosmology and astrophysics offer a laboratory of sorts to explore the effects of particle physics, \cite{dolgov1981cosmology}. 

In the standard cosmological model the source of the gravitational field is a perfect fluid and dark matter is often modelled as a pressureless and collisionless weakly interacting massive particle (WIMP) which in the light of experimental data (or rather lack of thereof) will need to be replaced with some other candidate \cite{PhysRevD.89.103526}. However, it is well known that the standard Cold Dark Matter (CDM) model seems to be inadequate on sub-galactic scales ($<1 $ Mpc) \cite{boehm2002interacting}.  This has motivated investigations into modifications to the standard model such as Warm Dark Matter (WDM) which damps the production of low-mass objects, \cite{schaeffer1988cold}. Alternatively, interactions in the dark sector have been investigated in recent years. Models of DM interactions include annihilation, scattering or decay (DDM) and offer simple extensions of the standard WIMP models which retain the desirable properties of CDM while offering possible resolutions to known tensions and discrepancies between theory and observation, \cite{mack2014known, boehm2002interacting, wang2014cosmological}. Strongly self-interacting modifications to CDM include collisional DM suggested by \cite{spergel2000observational} to account for flat density profiles of galaxy cores but ruled out by \cite{arabadjis2002chandra, boehm2005constraints}; and self-annihilation with a quadratic dependence on density is prevalent in galaxy cores, \cite{mack2014known}, but must be carefully contrived to avoid complete self-annihilation in the early universe, \cite{craig2001structure}. Scattering processes such as DM-photon or DM-neutrino interactions have been shown to have similar effects to warm DM models \cite{boehm2002interacting} while also producing horizon-scale weak damping effects. 

The effect of DM physics on cosmological structures offers a way of using cosmological observations to gain insight into possible interactions in the dark sector. Such observables include high redshift data such as cosmic dark ages and cosmic microwave background (CMB), \cite{slatyer2017general}, or low-redshift data such as the effects of dark matter on galactic halos \cite{mack2014known}. However astrophysical signals such as, for example, DM annihilation \cite{2014MNRAS.439.2728M, 2016PhRvD..93j3004L} have been shown to be difficult to distinguish from signals produced by baryonic processes \cite{2016PhRvD..94l3008L,2018JCAP...07..060C} making direct detection of the products of annihilation at best inconclusive. Interactions could introduce  additional effects that might influence the overall evolution of the universe. These effects, such as decay and scattering, could lead to energy or momentum flux, or viscosity making the fluid description of dark matter imperfect \cite{ellis1973relativistic,wainwright_ellis_1997,ellis2012relativistic}.

The focus of the present paper is decaying dark matter (DDM) which has a linear density dependence and may decay into two- or many- bodies with massless decay by-products, and may feature kick or injection velocities of relativistic order, \cite{blackadder2016cosmological}. Hence DDM may produce standard model particles, photons or exotic smaller mass dark particles or dark radiation, \cite{pandey2020alleviating, ackerman2009dark}, and energy-injection by cascade effects, \cite{slatyer2017general}, offering possible means of indirect detection. Unstable DDM is a natural development beyond the canonical stable WIMP model. 

One of the primary benefits of DDM is the early-time correspondence of the model to CDM and consistency with the CMB anisotropy spectrum  while late-time tensions within the CDM framework such as the missing satellite problem are offered potential resolutions by the long life-time instability of the DM particle, \cite{wang2014cosmological}, -- assuming a particle lifetime of order the current age of the universe or longer. In this scenario, the observed DM subhaloe underabundancy may be remedied by both decay into relativistic, \cite{cen2001decaying}, or non-relativistic daughter particles, \cite{wang2014cosmological}. Similarly, the correspondence of DDM with CDM at high-redshift offers possible alleviation to, if not resolution of, the observed tensions in $H_{0}$ as inferred from CMB and Type-1 Supernovae data, \cite{vattis2019dark}, as well as the $\sigma_{8}$ matter fluctuations, \cite{pandey2020alleviating}. However, recent work on relativistic single-species decay products have investigated the similarity of the induced bulk viscosity and modified cosmic acceleration with the negative pressure required of dark energy, \cite{wilson2007bulk}, and the signals constraining DDM physics may be degenerate to other model modifications such a modified Newtonian dynamics, \cite{peter2010constraints}. 

The present work proposes to investigate the possible decay of DM into other dark particles via the effect this may have upon the growth of large-scale structure, specifically cosmological voids. Voids are chosen as these offer an unique environment with several efficacious features. The under-abundance of galaxies in these avoided regions suggests an inhomogeneous DM under-density, \cite{peebles2001void}, assuming a strong Light-to-Mass ratio, and therefore cosmic voids will have less significant annihilation or self-interaction effects due to the quadratic mass dependency of these. Furthermore, these regions have minimal baryonic contamination so that scattering and baryonic feedback processes may be negligible. Thus, in cosmic void regions the only remaining particle interaction which may be introduced to modify DM and produce significant signals, is decay. 

As voids are the complementary structure of the filaments and sheets of the cosmic web, realistic analysis cannot consider voids as isolated objects, \cite{van2014voids}. The environmental dependence and hierarchical structure formation of voids creates a complex morphology which may conveniently be classified by the dominant evolutionary process determined by the surrounds: either expansion or collapse, \cite{sheth2004hierarchy}. Larger voids are observed to generally fit the former scenario corresponding to a void-in-void environmental configuration. The density profile of these voids rises smoothly towards the mean background density and are classified as R-type in recent literature. Alternatively, the density profiles of smaller, S-type voids exhibit a compensating shell suggesting a void-in-cloud type configuration, \cite{ceccarelli2013clues}. In both the uncompensated and compensated cases, there is a general trend of spherical voids evolving towards a bucket-shape density profile, \cite{van2014voids}. Furthermore, the smoothing of asphericities as under-densities grow, \cite{icke1984voids}, validates the approximation of spherically symmetric voids. Though not explored here, the ellipticity of voids is an important measure of tidal effects induced by the surrounding environment and offers an significant probe into cosmology and dark sector physics, \cite{rezaei2019effects, rezaei2020dark}. Despite the intimate coupling of voids with one another and the overall environment, the standard first-approximation of an isolated, spherically symmetric under-density is often adopted; we also do this here since it offers an immediate quantitative and qualitative appreciation of the effect of proposed dark matter decay upon void evolution. 

As these cosmological voids are inhomogeneous and evolve non-linearly, \cite{bolejko2006radiation}, the cosmology must move beyond the standard spatially homogeneous and isotropic Friedmann-Lema\^{i}tre-Robertson-Walker (FLRW) model. Perturbative approaches may be employed but require a stable background so cosmological back-reaction may affect the accuracy of these models, and furthermore, gauge issues may arise \cite{ellis1989covariant}. The Lema\^{i}tre-Tolman-Bondi (LTB) class of exact spherically symmetric dust solutions to the Einstein equations are inhomogeneous but isotropic and are appropriate for simplified void models as pressure is generally negligible until shell-crossing occurs, \cite{Lemaitre,LemaitreReprint,Tolman,TolmanReprint,1947MNRAS.107..410B}. The Szekeres models are both inhomogeneous and anisotropic, \cite{icm1997} and have been employed in the modelling of cosmological voids. These methods are however constrained by the requirement of being metric based. Recently, tetra-based methods have been developed as an alternative to LTB models, \cite{kim2019alternative}. The well-established semi-tetrad approach of the hydrodynamical 1 + 3 covariant methodology has an intuitive fluid flow interpretation, \cite{ellis1973relativistic}, and when taken to have local rotational symmetry (LRS) or silence (vanishing magnetic Weyl and pressure gradients) offers a generalisation to th LTB models in certain limits \cite{van1996covariant, van1997integrability}. In this paper we adopt a fully general relativistic hydrodynamical 1 + 3 formulation with LRS, to which further simplifications are imposed, see Section \ref{sec:methods}.

The decay effect is proposed to be modeled by the one-way transfer of energy-density via long-lifetime nuclear-like decay from a dust fluid comoving with the fundamental observer (representative of the CDM, with baryonic matter deemed insignificant and disregarded for the purposes of this model) into a secondary non-comoving fluid. This is intended to represent a simple exponential decay of the massive parent particle into a massive daughter particle. 

The average relative spatial velocity of the second non-comoving dust generated by the decay induces an effective imperfect fluid EM tensor in the frame of the fundamental observer. Tilted models wherein imperfect fluid effects arise have been studied in the literature, see \eg \cite{coley1990bianchi}. The presence of these imperfect terms corresponding to heat flux, pressure and viscosity induces non-FLRW effects such as shear or Weyl curvature (which are identically zero for the FLRW models), which in turn modifies the overall evolution of the cosmology and structure therein. However, for small peculiar velocities the EM tensor approaches that of a dust with heat conduction, investigated earlier in \cite{goode1986spatially} where an explicit form for the space-time metric was determined, and more recently in \cite{sebastian2021non}, wherein the momentum-transfer was interpreted as the effect of a non-comoving fluid with non-relativistic peculiar velocity, in accordance with the present work. 

The structure of this paper is as follows. In Section \ref{sec:methods}, the tilted two-fluid, 1 + 3 covariant model is developed which leads to a governing system of PDEs for a set of scalars once LRS is imposed. This system of equations is then employed in Section \ref{sec:results} to model spherically symmetric, isolated voids of the two types discussed above. The results of numerical analysis of this system are presented and discussed, with an emphasis on the possibility of using these results as an indirect means of constraining DDM. Limitations of our model are also discussed, and the article then concludes in Section \ref{sec:conclusion}.

\section{Methods}\label{sec:methods}

\subsection{The 1+3 covariant framework}

The tilted 1 + 3 covariant Lagrangian formalism is employed. The model is chosen primarily owing to its direct physical interpretation of the kinematic variables adopted, and, secondly, for its proven capability of addressing non-linear evolution of which voids are a weak example, \cite{maartens1999cosmic}. We also remark on its prevalence in the literature of relativistic cosmology. 

Following \eg \cite{ellis1973relativistic,ellis1999cosmological,tsagas2008}, flow-orthogonal projections are made to the spacetime by projection into the instantaneous rest space of an observer using
\[h^{a b} = g^{a b} + u^{a} u^{b}\] 
where $g^{a b}$ is the metric tensor and $u^{a}$ the comoving 4-velocity. The gradient of the 4-velocity field of the comoving observer is decomposed into vorticity, $\omega_{a b}$, shear $ \sigma_{a b}$, expansion $\Theta$ and acceleration $A_{a}$ such that
\begin{equation}
  u_{a ; b} = \omega_{a b} + \sigma_{a b} + \dfrac{1}{3} \Theta h_{a b} - A_{a} n_{b}  \, \text{.}
\end{equation}

As mentioned in the introduction, some DDM models postulate the production of standard matter particles which may be directly detected. However, in the present investigation, we wish to determine possible means of indirect detection of DDM via the imprint on cosmological voids, if any. Thus, for the purposes of this preliminary work, it is sufficient to assume that the secondary fluid produced via the decay is another dark fluid. 

This work proposes a model involving two dust components: the first is comoving with the fundamental observer, has density $\rho$ and a normalised comoving 4-velocity field $u^{a} = \delta^{a}_{t}$, and decays with a long-life time into the second non-comoving resultant fluid with density $\epsilon$ and 4-velocity $v^{a}$. The energy-momentum tensor for the total fluid is taken to be the sum of these dust components such that $T^{a b} = \Sigma_{(i)} T^{a b}_{(i)}$ and hence,
\begin{equation}
    T^{a b} = \rho u^{a} u^{b} + \epsilon v^{a} v^{b} \,\text{.}
\end{equation}

The non-comoving velocity is tilted with respect to the first by the Boost
\begin{equation}
\label{eq:boost}
    v^{a} = \gamma (u^{a} + V^{a}) 
\end{equation}
with Lorentz factor $\gamma = (1 - V_{a} V^{a})^{-1/2}$ and spacelike peculiar velocity $V^{a}$, \cite{king1973tilted}.

Hence, an observer moving with $u^a$, \ie comoving with the first fluid with density $\rho$, measures the total density to be
\[\mu = T^{a b} u_{a} u_{b} = \rho + \eta, \]
where $\eta = \gamma^2 \epsilon$, the total pressure to be 
\[p = \frac{1}{3} h_{a b} T^{a b} = \frac{1}{3} \eta h_{a b} V^{a} V^{b},\] the heat flux to be 
\[ q^{a} = -(\mu u^{a} + u_{a} T^{a b}) = \eta V^{a}, \]
and the anisotropic stress as
\[ \pi_{a b} = T_{c d} h^{c}_{\langle a} h^{d}_{b \rangle} . \] 

Thus, a comoving observer does not discern two dust components, but rather the source of the gravitational field is apparently a single imperfect fluid, which may be described by
\begin{equation}
     T^{a b} = \mu u^{a} u^{b} + p h^{a b} + 2 q^{(a}u^{b)} + \pi^{a b}.
\end{equation}

The assumption of LRS takes there to exist locally an axis of symmetry or preferred spatial direction induced by some properties of the flow (\eg shear or vorticity), such that at each event the spacetime exhibits partial isotropy, \cite{goode1986characterization}. Observations become independent of spatial dimensions orthogonal to the preferred spatial direction, \cite{2016PhRvD..94j4040S,2018arXiv180505664S}. Hence, all spacelike covariant vector fields in the LRS spacetime are proportional to the preferred spacelike unit direction vector. Let this preferred spacelike unit vector be denoted by $z^a$ and note that
\[  z^{a}z_{a} = 1, \quad  u^{a} z_{a} = 0. \] 

Two new variables may be introduced to describe the evolution of $z^{a}$. The magnitude of the divergence $\alpha$ is taken to be 
\[\alpha = D_{a} z^{a},\] 
where $D^{a} X = h^{a b} X_{; b}$ is the projected covariant derivative. The magnitude of the spatial rotation, or twist, $\tau$, is determined by 
\[\tau = - z_{a} \epsilon^{a b c d} u_{d} z_{c ; b},\]
where $\epsilon^{a b c d}$ is the Levi-Cita symbol. Note that it follows that, \cite{van1996covariant}, 
\[\dot{z}^{a} = 0, \quad z^{a}_{; a} = \alpha + A^{a} z_{a} , \] where $\dot{X} = u^{a} X_{; a}$. 

From the preferred spacelike unit vector $z^{a}$ is defined the symmetric tensor 
\[e_{a b} = \frac{1}{2}(3 z_{a} z_{b} - h_{a b}), \] 
which has the properties, \cite{van1996covariant}, that
\[\dot{e}_{a b} = 0, \quad D_{b} e^{a b} = \frac{3}{2} \alpha z^{a}, \quad  e^{a}_{b} e^{b}_{a} = \frac{3}{2}.\]

By LRS, in order to reduce the above equations to a scalar system, following \cite{van1996covariant} and \cite{bolejko2017relativistic}, all spacelike covariant vector fields may be represented as
\[ A^{a} = A z^{a}, \quad V^{a} = V z^{a}, \quad \omega^{a} = \omega z^{a} \]
from which it follows that 
\[ q^a = Q z^a, \quad \pi^{a b} = \Pi e^{a b}\]
and
\[ p = \frac{1}{3} \eta V^{2}, \quad Q = \eta V, \quad \Pi = \frac{2}{3} \eta V^{2}.\]  

In a similar fashion as above, LRS allows all the irreducible kinematic components of the shear and Electric and Magnetic parts of the Weyl tensors to be written in terms of scalars such as
\[ \sigma^{a b} = \Sigma e^{a b}, \quad E^{a b} = W e^{a b}, \quad H^{a b} = B e^{a b}.\]

The substitution of the above into the 1 + 3 covariant governing equations (see \eg \cite{tsagas2008}, and, adopting, for this section, units such that $c = 8 \pi G$), produces the following system of propagation equations:
\begin{align}
&\dot{\Theta} = -\dfrac{1}{3} \Theta^{2} - \dfrac{1}{2} (\mu + 3 p) - \frac{3}{2} \Sigma^2 \nonumber \\
& \quad \quad + 3 \omega^{2} + A^{'} + \alpha A + A^{2}  + \Lambda	\,\text{,} \label{eq:raych_gen} \\
&\dot{\Sigma} = - \dfrac{2}{3} \Theta \Sigma - \frac{1}{2} \Sigma^{2} - \frac{2}{3} \omega^{2} - W + \frac{1}{2} \Pi +\frac{2}{3} A^{'} - \frac{1}{3} A \alpha + A^{2}	\,\text{,} \\
&\dot{\omega} = - \frac{2}{3} \Theta \omega + \Sigma \omega + \tau A 	\,\text{,}  \\
&\dot{W} + \frac{1}{2} \dot{\Pi}  =  - \Theta W - \dfrac{1}{2} (\mu + p) \Sigma - \frac{1}{6} \Theta \Pi 
+ \frac{3}{2} \Sigma W  \nonumber \\
& \quad \quad - \frac{1}{4} \Sigma \Pi + \frac{1}{6} Q \alpha - \frac{1}{3} Q^{'} 	\,\text{,}     \label{eq:scalar_weyl_gen} \\
&\dot{B} = -\Theta B + \frac{3}{2} \tau W - \frac{3}{4} \tau \Pi + \frac{3}{2} B \Sigma - \omega Q 	\,\text{,}  \\
&\dot{\mu} = -\Theta (\mu + p) - \frac{3}{2} \Pi \Sigma - Q^{'} - Q \alpha - 2 Q A 	\,\text{,}  \\
&\dot{Q} = - (\Sigma + \frac{4}{3} \Theta) Q - \frac{3}{2} \Pi \alpha - \Pi^{'} - p^{'} - \Pi A - (\mu + p) A	\,\text{,}
\end{align}
where $X' = z^{a} X_{; a}$.  Moreover, we acquire the constraints:
\begin{align}
&Q = \dfrac{2}{3} \Theta^{'} - \Sigma^{'} - \frac{3}{2} \alpha \Sigma  - \tau \omega	\,\text{,} \label{eq:q_constraint} \\
&B = - \frac{3}{2} \tau \Sigma + \omega (2 A - \alpha) \,\text{,} \label{eq:b_constraint}  \\
&W^{'} = - \frac{3}{2} W \alpha + \frac{1}{3} \mu^{'} - \frac{1}{2} \Pi^{'} - \frac{3}{4} \Pi \alpha - \frac{1}{3} \Theta Q \nonumber \\ 
& \quad \quad + \frac{1}{2} \Sigma Q - 3 B \omega	\,\text{,} \label{eq:w_constraint} \\
&B^{'} = - \frac{3}{2} \alpha B + (\mu + p) \omega + \frac{1}{2} \tau Q + 3 \omega W - \frac{1}{2} \Pi \omega \,\text{,}  \\
&\omega^{'} = (A - \alpha) \omega  \,\text{.} 
\end{align}

\subsection{Pressureless, irrotational and dissipative fluids}

In order to further simplify the system the stringent conditions of irrotationality \ie vanishing vorticity, $\omega = 0$, and twist, $\tau = 0$, are assumed. The former ensures the existence of a well-defined cosmological time. It follows via the constraint equation (\ref{eq:b_constraint}) that a twist-free, irrotational flow is necessarily purely gravito-electric such that $B = 0$.

The vanishing of the effective pressure scalars $p$ and $\Pi$ and heat scalar $Q$ immediately follows in the limit of $V \to 0$. In this situation, the model reduces to the well-know spherically symmetric LTB cosmology. As mentioned in the introduction, exact homogeneous solutions featuring dissipative dust were found and analysed in \cite{goode1986spatially}. Early research explored inhomogeneous dust models with heat flow, generalising the subclasses of inhomogeneous and anisotropic Szekeres solutions \cite{de1985chaotic}. However, as recently pointed out by \cite{najera2020non}, the heat flow of dissipative dust models is inappropriate for standard CDM and late-time universe models, and is analogous to the energy flux induced in the observer frame by a non-comoving, non-relativistic peculiar velocity. The two-fluid tilted model developed here has produced an effectively imperfect total EM tensor.  

Observational analyses of large-scale peculiar velocities suggest non-relativistic orders. It therefore seems valid to assume a non-relativistic, non-comoving peculiar velocity of the second fluid such that $\gamma \approx 1$, and the effective pressure scalars $p$ and $\Pi$ become second-order and negligible. The effective heat scalar $Q$, however, is not. The tractability of the mathematical model is further increased by assuming that the covariantly projected dot and prime derivatives of the effective pressures are also first order and negligible. This is justified as the initial velocity distribution is taken to be homogeneous, and hence, all terms featuring $p$ or $\Pi$ vanish. 

The above assumptions lead to the set of propagation equations:
\begin{align}
&\dot{\Theta} = -\dfrac{1}{3} \Theta^{2} - \dfrac{1}{2} \mu - \frac{3}{2} \Sigma^2 + A \alpha + A^2 + \Lambda	\,\text{,} \label{eq:raych_eq} \\
&\dot{\Sigma} = - \dfrac{2}{3} \Theta \Sigma - \frac{1}{2} \Sigma^{2} - W - \frac{1}{3} A \alpha + \frac{2}{3} A^{2} \,\text{,} \label{eq:scalar_shear} \\
&\dot{W}  =  - \Theta W - \dfrac{1}{2} \mu  \Sigma + \frac{3}{2} \Sigma W + \frac{1}{6} Q \alpha  - \frac{1}{3} Q'  - \frac{2}{3} A Q	\,\text{,} \label{eq:scalar_weyl} \\
&\dot{\mu} = -\Theta \mu  - Q' - Q \alpha - 2 A Q	\,\text{,}  \label{eq:scalar_rho} \\
&\dot{Q} = - (\Sigma + \frac{4}{3} \Theta) Q - \mu A	\,\text{,} \label{eq:heat_prop}
\end{align}
and the constraints:
\begin{align}
&Q = \dfrac{2}{3} \Theta' - \Sigma' - \frac{3}{2} \alpha \Sigma 	\,\text{,} \\
&W^{'} = - \frac{3}{2} W \alpha + \frac{1}{3} \mu' - \frac{1}{3} \Theta Q + \frac{1}{2} \Sigma Q 	\,\text{.} 
\end{align}

Therefore, to close this system an evolution equation for the unknown spatial divergence $\alpha$ is required. This is determined by application of the Voss-Weyl divergence formula to the comoving 4-velocity $u^{a}$, and to the preferred spatial direction $z^{a}$, giving, respectively
\begin{equation}
    \Theta = \frac{\dot{\nu}}{\nu} \quad \text{ and} \quad \alpha = \frac{\nu'}{\nu} \,\text{,} \label{eq:nu_eqs}
\end{equation} 
where $\nu = \sqrt{-g}$, with $g$ representing the metric determinant.

Hence, the system formed by Equations (\ref{eq:raych_eq}) - (\ref{eq:nu_eqs}),  serves as a closed set of evolution equations which, given appropriate initial conditions, may be numerically marched forward in time. The constraints may be helpful as an alternate substantiation of the results of evolving the system. 

In addition, one can define the scale factor $a$
\begin{equation}
  a = \left( \frac{\nu}{\nu_i} \right)^{1/3}. 
  \label{scalefactor}
\end{equation}
In general the scale factor defined in this way is a function of both time and position but in the FLRW limit the scale factor becomes a function of time only, \ie $a=a(t)$ and $a' = 0$.

\subsection{FLRW limit}\label{FLRW}

In order to facilitate a comparison with the standard model and a discussion of the modifications introduced, briefly consider necessary and sufficient conditions for a spacetime to be FLRW \cite{icm1997}:
\begin{enumerate}
    \item the metric obeys the Einstein equations with a perfect fluid source; and
    \item the velocity field of the perfect fluid source has zero rotation, shear and acceleration. 
\end{enumerate}

The first condition requires
\[ Q = 0,  \]
while the second condition gives
\[ \Sigma = 0, {\rm ~~~and~~~} A = 0. \]

Since these requirements must be fulfilled at all times, it follows from (\ref{eq:scalar_shear}) that 
\[W = 0 .\]
This is expected as the FLRW models are conformally flat (\ie the Weyl curvature vanishes). These conditions, together with (\ref{eq:w_constraint}) require that $\mu'=0$, which in turn via (\ref{eq:scalar_rho}) implies that $\Theta'=0$, meaning all gradients vanish, including $a'=0$.
 
Using (\ref{eq:nu_eqs}) and (\ref{scalefactor}) it follows that the Hubble parameter
\begin{equation}
 \frac{\dot{a}}{a} \equiv H = 3 \Theta.
 \label{Hubble}
\end{equation} 
This implies that the Raychaudhuri equation (\ref{eq:raych_eq}) reduces to the 2nd Friedmann equation, which together with the continuity equation (\ref{eq:scalar_rho}) yields the 1st Friedmann equation
\begin{equation} \label{eq:F1}
\frac{\dot a^{2}}{a^2} =  - \frac{Kc^2}{a^2} + \frac{8\pi G}{3} \rho + \frac{1}{3} \Lambda c^2,
\end{equation}
which is usually conveniently rewritten at the present instant ($t=t_0$) in terms of the $\Omega-$parameters
\begin{equation} \label{eq:omega_sum}
\Omega_M + \Omega_\Lambda + \Omega_K = 1,
\end{equation}
where
\begin{align}
& \Omega_K = -\frac{K}{H_0^2 a_0^2}, \label{Omegak}\\
& \Omega_M = \frac{8 \pi G}{3 H_0^2 } \, \rho_0, \label{Omegam} 
\end{align}
and
\begin{align}
&  \Omega_\Lambda =  \frac{1}{3 H_0^2} \, \Lambda.  \label{Omegal} \end{align}
Cosmological parameters determined by the 2015 \textit{Planck} observations are adopted so that $\Omega_K = 0$, $\Omega_M = 0.308$,  $\Omega_\Lambda = 0.692$ and $H_0 = 67.81$ km s$^{-1}$ Mpc$^{-1}$, \cite{2016A&A...594A..13P}.

\subsection{Initial conditions} 
 
The model begins from the last scattering instant (CMB), the redshift of which is taken to be $z_{CMB} = 1090$. At the initial instant, it is assumed there exists only a single, comoving dust fluid. Then $V = 0$ and the system is approximated by an exact spherically symmetric solution of the Einstein equations for the gravitational field sourced by dust only, \ie the LTB model. In this, the fluid quantities are
\begin{eqnarray}
&& \Theta = \frac{ \dot{R}'}{ R'} + 2 \frac{\dot{R}}{R}, \label{ltbtht} \\
&& \Sigma = -\frac{1}{3} \left(  \frac{ \dot{R}'}{R'}  - \frac{\dot{R}}{R} \right),  \label{ltbshr} \\
&& {\cal W} = \frac{1}{3} \left( \frac{M}{R^3} -\frac{M'}{R^2R'}\right), \label{ltbwey} \\
&& \mu =  \frac {2 M' } {R^2 R'}, \label{ltbrho} \\
&& Q = 0. \label{ltbheat}
\end{eqnarray}

The evolution equations can be reduced to a single ODE \cite{2009suem.book.....B}
\begin{equation}\label{evo}
\dot{R}^2 = -K(r) + \frac {2 M(r)} {R} + \frac 1 3 \Lambda R^2.
\end{equation}
Thus, to specify the LTB model, and in our case the initial condition, one is required to provide the two functions: $M(R)$ and $\dot{R}(R)$.

The two void configurations discussed in the introduction will be explored in this work by specifying initial conditions for these types separately. In both cases it is assumed that the radial coordinate coincides initially with the function $R$, \ie
\begin{align} \label{initial}
 &   r = \left. R \right|_{t_i} ~~\Rightarrow~~ \left. R' \right|_{t_i}= 1 \\
&    t_i = t_{CMB}.
\end{align}

The above is equivalent to defining the scale factor to be $a_i=1$ when $t = t_{i}$. The standard convention is to set the scale factor at present to unity and put $a(z) = 1/(1+z_{CMB})$. However, this scaling does not have any physical meaning, so does not influence the evolution of the system, and it is more convenient to set the scale factor to unity at the CMB (when the model is almost homogeneous) than at the present time (when the system is far from FLRW). 

Due to the system being not FLRW the evolution of the radius $R$ does not follow a simple FLRW scaling so that $R\ne ar = r (1+z)/(1+z_{CMB})$. However, for the structures considered here (cosmic voids and present-day scales of Mpc), when calculating the physical distance from the origin to a considered point, this is approximated by
\begin{equation}
 R(r,t) \approx \int\limits_0^r \, {\rm dr } \, a(t,r),    
 \label{radius}
\end{equation}
where the value of the scale factor $a(t,r)$ follows from (\ref{scalefactor}). The above approximation neglects the contribution from the spatial curvature $K$. While the function $K$ is known at the initial instant, dissipate processes mean that it is not a function of $r$ (as in LTB models) but evolves in time \cite{2011CQGra..28p4002B}.

It is important to emphasise that the value of $R$ does not affect the evolution of the system governed by (\ref{eq:raych_eq}) -- (\ref{eq:heat_prop}). Rather, it is only used for plotting purposes to represent graphically the physical quantities, \ie rather than plotting fluid variable (\ie $\mu, \Theta, \Sigma, W$, or $Q$) as a function of r (in Kpc), in Sec. \ref{sec:results} these quantities are plotted in terms of $R$, as given by (\ref{radius}).

    \begin{figure*}
    \begin{center}
      \includegraphics[scale=0.52]{figure1.eps}
    \end{center}
    \caption{Density profile at $z = 0$ of the evolved S-type void with (blue solid lines) and without (red dashed lines) decay, for various combinations of the parameters $\Gamma$ and $v_i$. This panes clearly illustrate the injection-velocity-dependent growth of novel secondary structure at the edge of the under-dense region.}
    \label{fig1}
    \end{figure*}

\subsubsection{S-type voids}

The S-type voids are surrounded by a compensating shell, \ie the mass of the system outside the void is approximately the same as the mass of a region of a similar size but otherwise homogeneous. For the S-type voids, the initial profile for the function $M$ is taken to be 
\begin{equation}
M(r) = \frac{1}{6}  \rho_{CMB} \left[ 1 + \frac{1}{2} m_0 \left( 1 - \tanh \frac{r-r_0}{2 \Delta r} \right)  \right] r^3, 
\end{equation} 
with $m_0 = -0.006$, $ \Delta r = 1.3$ [Kpc], and $r_0 = 6.0$ [Kpc], $\rho_{CMB} = (1+z_{CMB})^3 \Omega_M 3 H_0^2/(8 \pi G)$ and $z_{CMB} = 1090.0$. The function $K$ is fixed by the condition of a uniform age of the universe $t_B = 0$, so that it corresponds to existence of pure growing modes \cite{2009suem.book.....B,2013CQGra..30w5001S}. This set of parameters leads to a void that at its central part has a density contrast of $\delta_0 = -0.9$ (cf. Fig. \ref{fig1}). 

\subsubsection{R-type}

\begin{figure*}
    \begin{center}
      \includegraphics[scale=0.52]{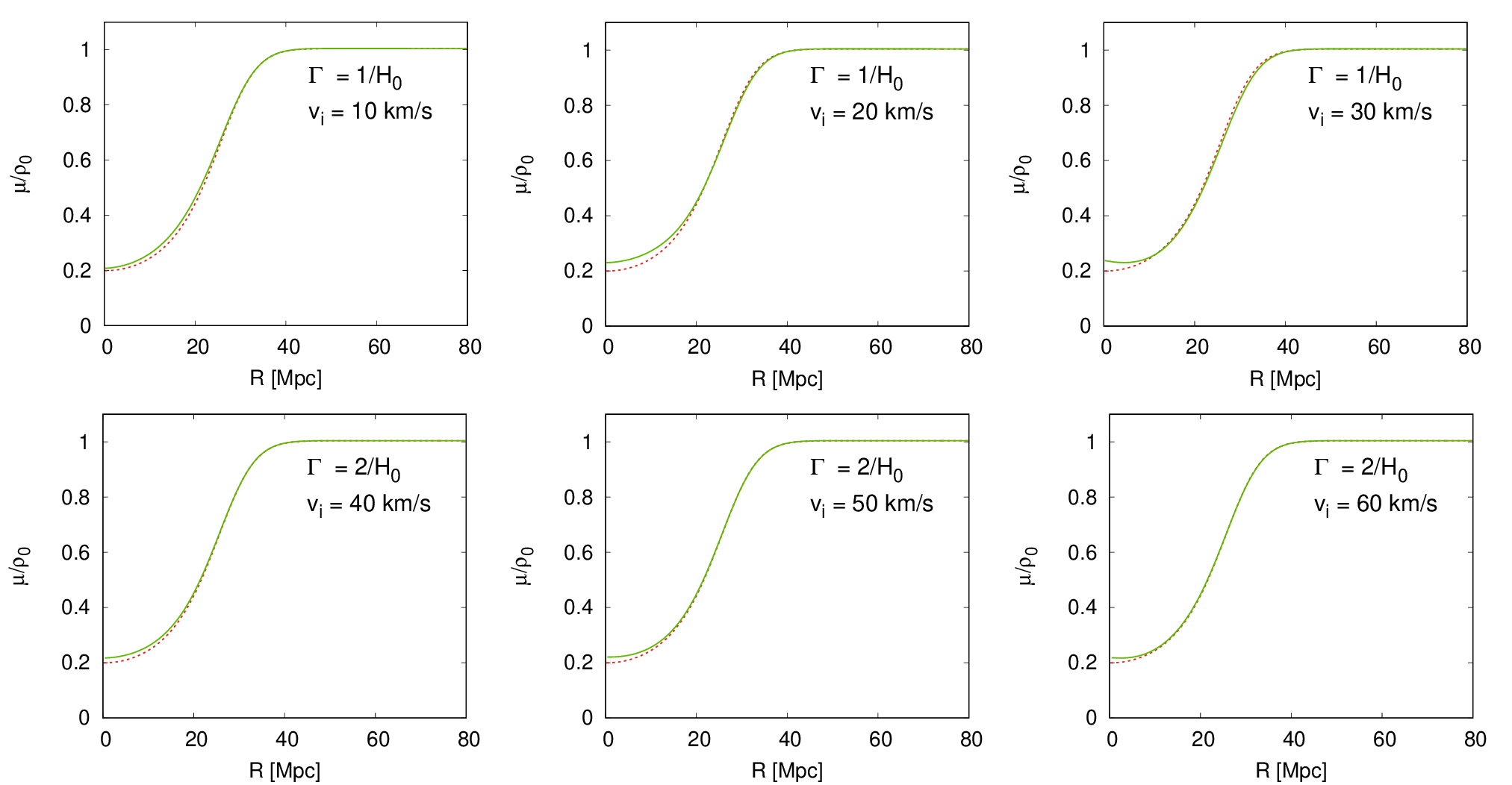}
    \end{center}
    \caption{Density profile at $z = 0$ of the evolved R-type void with (blue solid lines) and without (red dashed lines) decay, for various combinations of the parameters $\Gamma$ and $v_i$. Unlike in the S-type case, there is only apparent a slight shallowing of the centre of the R-type under-density.}
    \label{fig2}
    \end{figure*}

A void of R-type has a continuously rising density profile, without a compensating shell. Such a shape may be modelled with a pure Gaussian density fluctuation, and then it follows from (\ref{ltbrho}), that the function $M(r)$ when $t = t_{i}$, is
\begin{equation}
M(r) = \frac{1}{2} \rho_{CMB} \int\limits_0^r {\rm d} \tilde{r} \, \tilde{r}^2  \left[ 1 + m_0 \exp \left( -
\frac{\tilde{r}^2}{\sigma^2} \right) \right],
\end{equation}
where $ \rho_{CMB} $ is as above, and $m_0 = -0.0033$, $\sigma = 24$ [Kpc]. As above, the  function $K$ is fixed by the condition of pure growing modes perturbations \citep{2009suem.book.....B,2013CQGra..30w5001S}. This set of parameters leads to a void that at its central part has a density contrast of $\delta_0 = -0.8$ (cf. Fig. \ref{fig2}). 

    \begin{figure*}
    \begin{center}
      \includegraphics[scale=0.52]{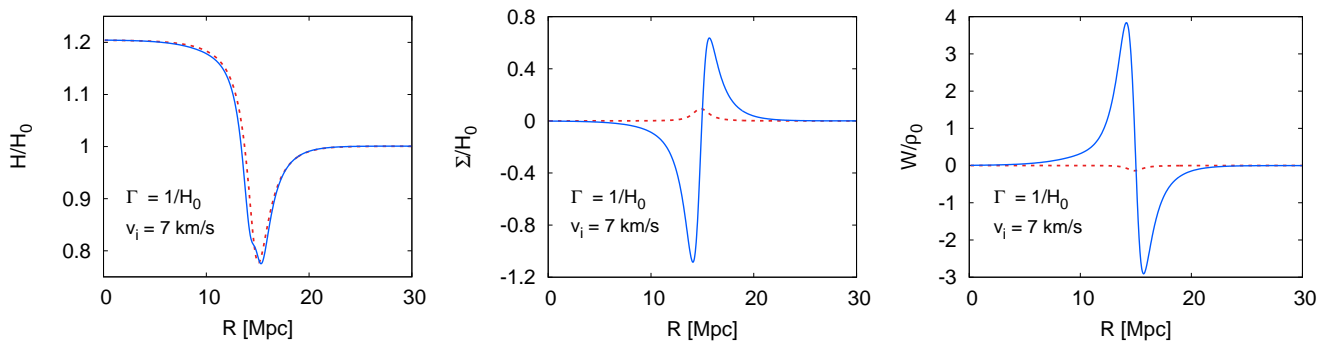}
    \end{center}
    \caption{Present day form of the non-dimensionalised $H$ (see eq. (\ref{Hubble})), shear scalar $\Sigma$, and electric Weyl scalar $W$ for the S-type void with parameters $\Gamma = 1/H_0$ and $v_i=7$ km/s. There is an insignificant difference in the scaled Hubble parameter between the decaying and non-decaying models, while significant changes in the latter two parameters.}
    \label{fig3}
    \end{figure*}

\subsection{Dark Matter physics} 

The gravitational field in the present model is sourced by DM only which is assumed to consist of two fluids: the original comoving dark dust fluid that decays into a secondary dark dust fluid which is non-comoving. In order to account for the interaction, the conservation equations for each fluid must be satisfied in the form 
\begin{equation}
    T^{a b}_{(i)\, ; b} = I^{a}_{(i)}
    \label{eq:conservation}
\end{equation}
where $ I^{a}_{(i)}$ is the interaction term such that the total fluid is conserved, $\Sigma_{(i)} I^{a}_{(i)} = 0$, \ie $I_{(2)} = - I_{(1)}$.

For this model of decay the interaction term is assumed to have the following form
\begin{equation}
I_{(1)}^{a} = - \Gamma \rho (u^a + w^{a}) \, \text{.}
\end{equation}
where $w^a = v_i z^{a}$ is a vector that describes initial velocity of the second fluid immediately after decay.

The temporal component of (\ref{eq:conservation}) is a  typical decay law. Indeed, using $\rho = N/V$ and  $\dot{V}/V = \Theta$, this equation can be re-written as 
\[ \dot{N} = -\Gamma N, \] 
which is a typical decay law of exponential (or nuclear) decay. The spatial component of (\ref{eq:conservation})
allows us to evaluate the acceleration $A$ since
\begin{equation}
\rho A - z_{a} I^{a}_{(1)} = 0 ~~\Rightarrow~~ A=  \Gamma v_i, \label{eq:acc_eq}
\end{equation}
where $v_i$ is the initial (or the injection or kick) velocity of the particles of the second fluid with respect to the decaying fluid. In practice $v_i$ is the average velocity of the `injected' particles. Therefore, this model contains two parameters: the decay rate $\Gamma$ and and injection velocity $v_{i}$.

    \begin{figure*}
    \begin{center}
      \includegraphics[scale=0.52]{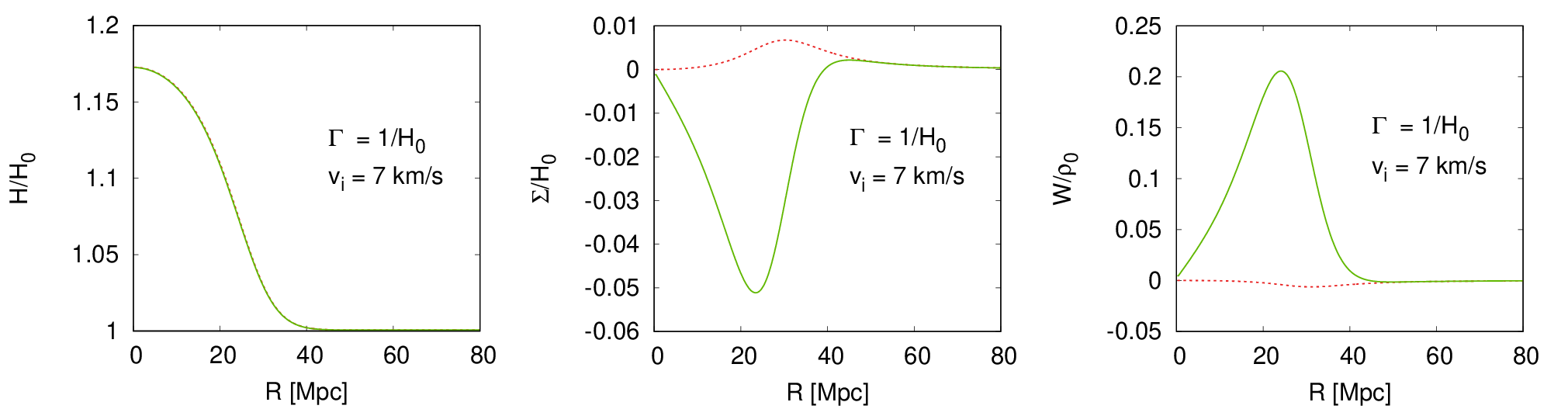}
    \end{center}
    \caption{Present-day form of the non-dimensionalised $H$ (see eq. (\ref{Hubble})), shear scalar $\Sigma$, and electric Weyl scalar $W$ for the R-type void with parameters $\Gamma = 1/H_0$ and $v_i=7$ km/s.}
    \label{fig4}
    \end{figure*}

        \begin{figure*}
    \begin{center}
      \includegraphics[scale=0.52]{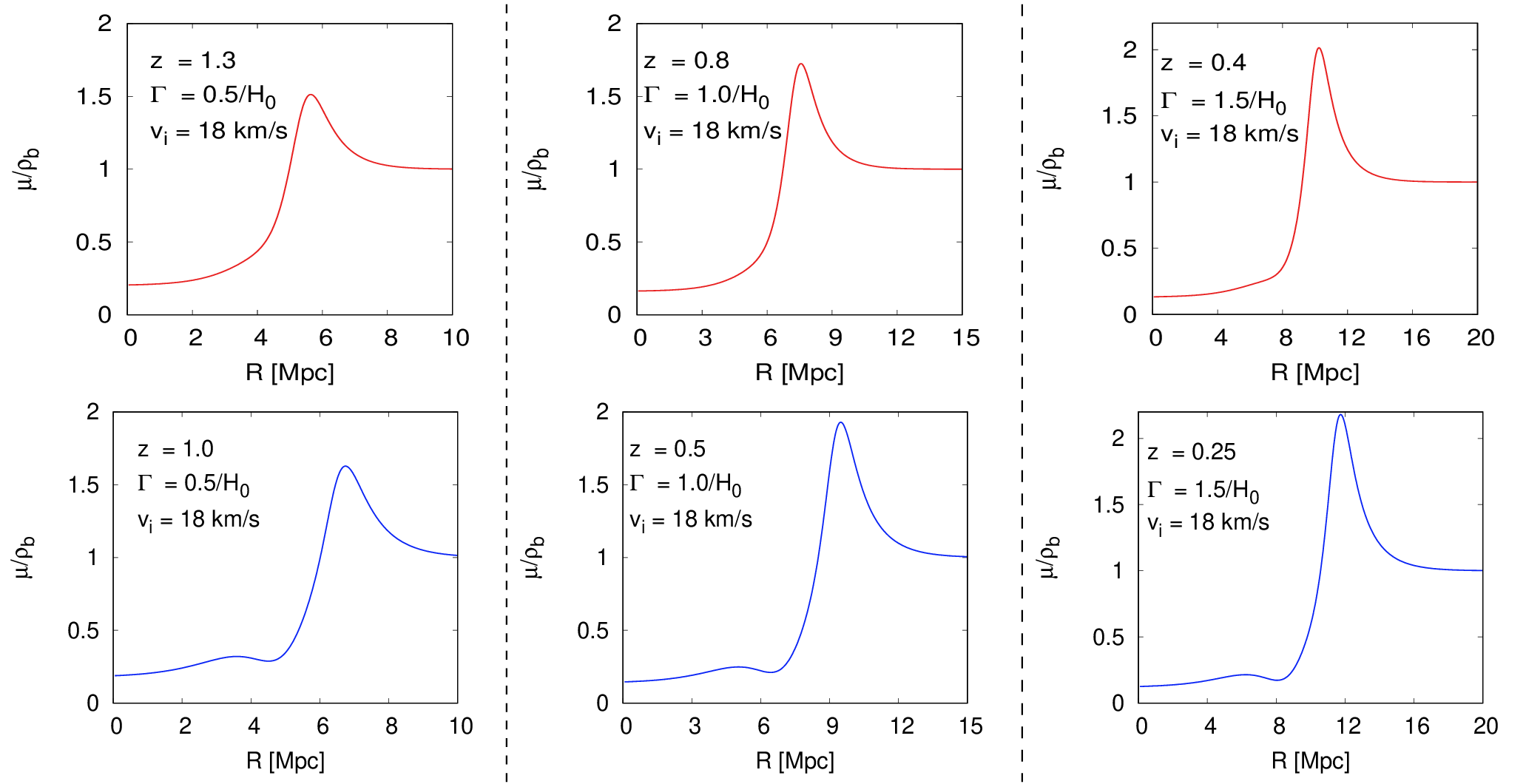}
    \end{center}
    \caption{Evolution of density profile for S-void at different redshift and for different values of the decay rate $\Gamma$. Top panels show the void without the feature present, while bottom panels show voids with the feature present, thus allowing to link the lowest redshift where no feature is present with the properties of dark matter.  Due to the degeneracy between $\Gamma$ and $v_i$ in order to use observations of voids at different redshifts to constrain the decay rate a further information about $v_i$ (or equivalently particle mass ratio and energy released at the decay) is required. }
    \label{fig5}
    \end{figure*}

Investigations of self-gravitating halos of nearly-degenerate, non-relativistic DDM have produced observational constraints that rule out the combination of decay lifetimes $\Gamma < H_{0}^{-1}$ Gyr and injection velocities of $v_{i} > 20$ km/s \cite{peter2010constraints}. Analysis of the observed mass-concentration and cluster mass function has constrained the DDM parameter-space of decay lifetimes and kick velocities for numerical simulations of DDM, suggesting an allowed region of lifetimes from a few times longer than the age of the universe to less than $100 H_{0}^{-1}$ and kick velocities $1 \, \text{km/s} < v_{k} < 100 \, \text{km/s}$, \cite{peter2010dark}. Other indirect constraints on the parameters of generic DDM models have been developed via the luminosity distance of supernovae standard candles which has a cosmological and energy-density dependence, \cite{blackadder2014dark}. These results suggest that in the case that 90\% of the mass is transferred to the massive daughter particle, the decay lifetime has a lower limit of $8.4$ Gyr with 95\% confidence. Alternatively, when 15\% is transferred, the lower limit on the lifetime is 25 Gyr, \cite{blackadder2016cosmological}. Other observational constraints on DDM models have been studied in the literature by means such as cosmological weak lensing effects \cite{wang2012effects} or the influence on reionisation via photons emitted during the decay process \cite{mapelli2006impact}. In view of this preceding research, we propose to investigate a range of decay parameters within similar regions. We investigate the region of parameter space relating to decay lifetimes of between one and a few times the age of the universe, and small non-relativistic injection velocities.

\section{Results}
\label{sec:results}
    
Starting from the initial conditions outlined above, the governing system of propagation equations, (\ref{eq:raych_eq}) -- (\ref{eq:heat_prop}), was time-marched forward to the present day, $z = 0$. The results of the evolution of the  S-type void is presented in Fig. \ref{fig1} and the R-type void in Fig. \ref{fig2}. The density profiles shown in these figures are normalised by the present day background density given by eq. (\ref{Omegam}).
Figure \ref{fig1} illustrates the  evolution of the S-type void over a range of decay-rates $\Gamma$ and injection velocities $v_i$. Overlaid in each plot as a red dashed line are the results of the evolution of the density profile without decay, $\Gamma = 0$, and thus no non-comoving second fluid nor effective heat/momentum flow (\ie such a model reduces to the LTB model).

The results summarised in Fig. \ref{fig1} indicate that the heat flow associated with the decay leads to a formation of a new feature within a void. This structure may be interpreted as an additional ditch-like underdensity adjacent to the edge of the void. The formation of this secondary structure can occur either in the case of a short decay rate (\ie the decay starts earlier in the past) and low injection velocity or, alternatively, with a longer decay rate (\ie decay starts at a later stage of evolution) and larger injection velocity. This is due to the fact that there is an interplay between the two parameters $\Gamma$ and $v_i$ wherein the decrease in one may be compensated by the increase of the other. This degeneracy follows from eq. (\ref{eq:acc_eq}), which shows that the two parameters are both encapsulated in the effect of the scalar acceleration term.

A selection of total density profiles is displayed in Figure \ref{fig2} for an uncompensated R-type profile and various parameters. In the case of the R-type voids, apart from a very slight shallowing of the centre (in comparison to the case without decay), there is no apparent impact on the mass distribution inside the void.
The reason why R-type voids do not exhibit distinctive features is related to the fact the dark matter decays effects are second-order, which is more clearly evident in Fig. \ref{fig3} and \ref{fig4}.
Initially (at the CMB instant), the models considered in this paper, are very close to the FLRW model, the evolution of which is driven by density and the expansion rate only (\cf Sec. \ref{FLRW}). Cosmic voids are formed due to the growth of perturbations imposed on the FLRW background \cite{2005MNRAS.362..213B,2006MNRAS.370..924B,2007PhRvD..75d3508B}. 
Perturbations in the expansion rate and density are first-order whereas, as seen from 
eqs. (\ref{eq:raych_eq}) -- (\ref{eq:heat_prop}), both shear $\Sigma$ and Weyl curvature $W$ are clearly second-order. 
Figures \ref{fig3} and \ref{fig4} show the present day form of the expansion rate, shear, and Weyl curvature. 
The expansion rate is expressed in terms of the Hubble parameter $H = \Theta / 3$.
For both types of voids (S-type and R-type) the perturbation in the expansion rate is large but still very similar to the LTB case (\ie the inhomogeneous model but without presence of heat flux). However, the largest difference between these models is seen in the plots of shear $\Sigma$ and Weyl curvature $W$. 
In the case of the S-type models the shear $\Sigma$ and Weyl curvature $W$  are clearly distinct  and large.
In contrast, while for the R-type models the shear $\Sigma$ and Weyl curvature $W$ are distinct from the LTB case they are now small, so their contribution to the evolution of the system is slight. 
The results presented in Figs. \ref{fig3} and \ref{fig4} indicate that the existence of the additional ditch-like underdensity forming next to the edge of the void is a second-order effect associated with decay-related heat flux. 
The process of forming this substructure is dynamical and interrelated: 
as seen from eqs. (\ref{eq:raych_eq}) -- (\ref{eq:heat_prop}), the density evolution is affected by the divergence of the heat flux $Q$, which in turn is influenced by shear $\Sigma$; the shear itself depends on $W$, which is a function the divergence of heat flux. Thus if the gradients are small, the growth of the heat flux and heat-flux related growth of shear and Weyl is limited -- this is the case of the R-type voids. 
On the other hand, in S-type voids the gradients are large and their presence enhances the growth of the shear and $W$, and leads to the formation of the additional ditch-like underdensity.

    \begin{figure}
    \begin{center}
      \includegraphics[scale=0.65]{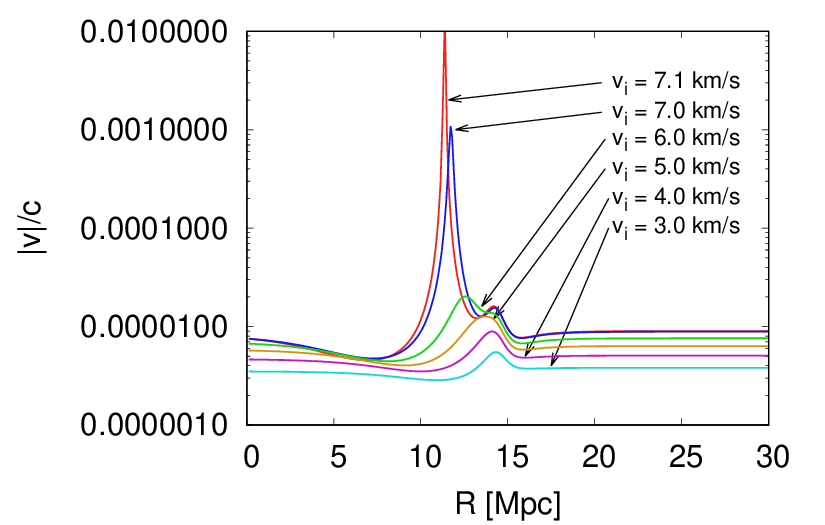}
    \end{center}
    \caption{Present day velocity of the second fluid, $v = Q/\eta$, for different values of the injection velocity $v_i$ and $\Gamma = 1/H_0$. For the S-type void, the approximation of small velocity breaks for $v_i > 7$ km/s. }
    \label{fig6}
    \end{figure}

\begin{figure}
    \begin{center}
      \includegraphics[scale=0.65]{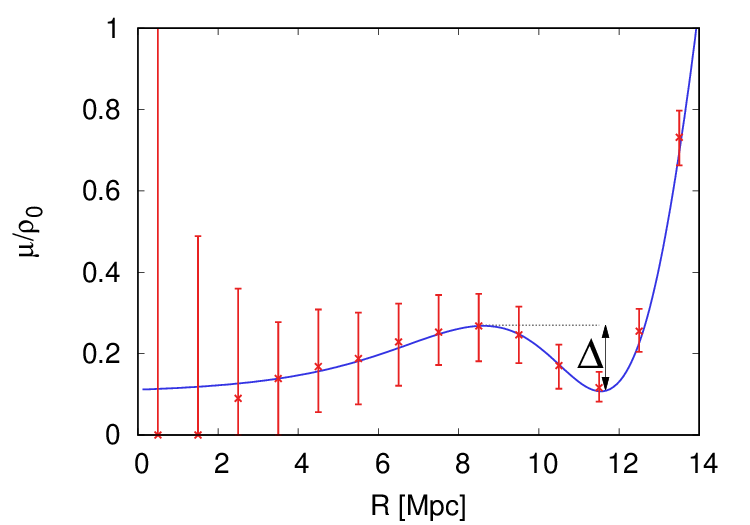}
    \end{center}
    \caption{Reconstruction of a density profile based on the number counts of tracer within a void. The solid line shows the present-day  density profile of an S-type void with $\Gamma = 1/H_0$, $v_i = 7$ km/s. The decrease in density near the edge of the void is $\Delta = 0.16$. The number of points required to obtain presented distribution is 1400 points. Assuming that there are approximately 5 galaxies per void, this corresponds to stacking of approximately 280 voids of similar size and spherical shape.}
    \label{fig7}
    \end{figure}

The extra sub-structure opens new avenues for using cosmic voids in studies of the physics of DM. Detection of such a feature would suggest the presence of heat flux associated with the decay of DM. In addition, one could target voids at different redshifts to detect when the secondary structure starts is initiated. Figure \ref{fig5} shows a  series of snap-shots of the evolution of an S-type density profile at a few different redshift and different decay rates $\Gamma$. As the models with shorter decay rate develop this new feature sooner, this structure will be apparent at a higher redshift. Thus by targeting voids at different redshift one could, in principle, attempt to constrain various particle properties such as the half-life $t_{1/2} = \Gamma^{-1} \ln(2)$, the average decay injection velocity $v_i$, or even particle masses. However, the results presented in Figure \ref{fig5}, should only be treated a proof-of-concept. 

First of all, as mentioned above, there is a degeneracy between the decay rate $\Gamma$ and injection velocity $v_i$. Figure \ref{fig5} shows snapshots for $v_i = 18$ km/s. With a longer decay rate, one could also obtain the feature at higher redshifts if the injection velocity is sufficiently large. Thus, without additional constraints on properties of dark matter particles (such as masses and energy released during the decay) one may not be able to disentangle constraints on $\Gamma$ from $v_i$. 

Secondly, and most importantly, the model considered in this paper is built on a framework that assumes that the relative velocity of one fluid with respect to another is small.  This neglects contributions from pressure and pressure gradients. Thus, the framework breaks once the velocity of one fluid with respect to another is sufficiently large.

In the linear regime of small perturbations, the relative velocity of fluids decays. It is only when second-order effects (such as shear and Weyl curvature) become non-negligible that the relative velocity starts to increase \cite{gaspar2019non}. Note that in the cases considered here, the decay-related heat flux is sensitive to the injection velocity. If the injection velocity is too large, the growth of the heat flux enhances the growth of the shear and $W$, which in turn leads to a rapid growth of the relative velocity and to a failure of the framework implemented in this paper.
This phenomenon is illustrated in Fig. \ref{fig6}. If the injection velocity is larger than  $v_i = 7.0$ km/s the model breaks at the present day. If the injection velocity is larger then the model fails earlier. 
For the examples presented in Fig. \ref{fig5}: the model with $\Gamma = 0.5 H_0^{-1}$ and $v_i = 18$ km/s fails approximately at $z = 0.8$, while the model with $\Gamma = H_0^{-1}$ and $v_i = 18$ km/s stops at $z = 0.4$.
While the results presented here open a new possibility of testing dark matter physics, \ie
by observing this secondary structure at certain red-shifts, the framework requires further development in order to quantitatively allow observational data to constrain properties of dark matter physics.

Such a structure could be observed either directly, by mapping the galaxy distribution inside cosmic voids, or, indirectly, by investigating light propagation through cosmic voids. 
Due to a limited number of galaxies inside cosmic voids (a few counts per void),
it is not possible to use the galaxy counts within a single void to detect this structure. Rather, one requires a large number of voids for stacking the signal. By stacking voids of similar size and shape, one can increase the number of tracers inside a ``stacked void". Such a procedure is presented in Fig. \ref{fig7} which shows an estimate of the 
(minimal) number of tracers needed to detect a decrease of number of galaxies near the edge of a void. 

The solid line in  Fig. \ref{fig7} shows the present-day  density profile of an S-type void with $\Gamma = 1/H_0$, $v_i = 7$ km/s. This density profile was used to generate a random distribution of tracers within a galaxy. The points represent the average density obtained by taking a ratio of galaxies $N$ within a given radius-bin to average number of galaxies $\bar{N}$ (\ie assuming galaxies are uniformly distributed, with number density $\bar{n}$, the expected number of galaxies within the radius-bin is $\bar{N} = \bar{n} V$, where $V$ is the volume of a particular distance-bin). The error bars represent 95\% confidence interval which was inferred from Monte Carlo simulations: using the density profile (solid line) random number of galaxies within a void was generated; the density then followed $\rho/\rho_0 = N/\bar{N}$; repeating the process large number of times and excluding lower 2.5\% and upper 2.5\% estimates, resulted in the 95\% confidence interval. The required number of tracers for the case presented in Fig. \ref{fig7} is 1400. So, assuming, 5 galaxies per void \cite{2012MNRAS.421..926P,2014MNRAS.442.3127S}, this means that at least approximately 300 voids of similar size and spherical shape would be required to detect the decrease in density $\Delta = 0.16$.

\begin{figure}
    \begin{center}
      \includegraphics[scale=0.65]{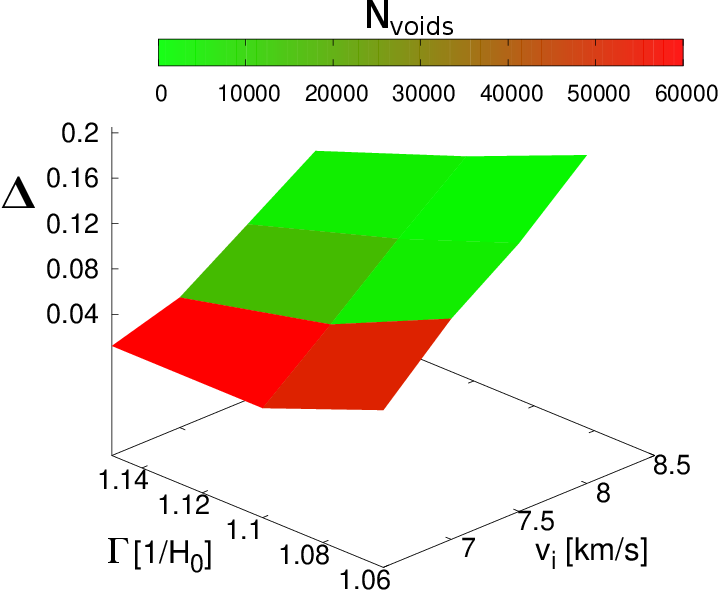}
    \end{center}
    \caption{
The fractional decrease in density  near the edge of the voids $\Delta$  as a function of DDM parameters $\Gamma$ and $v_i$. Colors correspond to a minimal number of voids required to detect such a feature. We emphasise this should be treated as a simple estimate obtained using a direct method based on the number counts of galaxies within cosmic voids. This was done by stacking voids of the same size and shape. Given that real voids are of different size and shape as well as are subject to intrinsic variations in density profiles, the actual number of required voids is most likely much higher.}
    \label{fig8}
    \end{figure}

The decrease $\Delta$ depends on the parameters of void and properties of DDM (such as $\Gamma$ and $v_i$). The larger the decay rate or the smaller the injection velocity, the smaller value of $\Delta$, and hence more voids would be required to detect this feature. This is demonstrated in Fig. \ref{fig8} which shows that when $\Gamma > 1.05 H_0^{-1}$ and $v_i < 7$ km/s the decrease is lower than $\Delta = 0.05$ and that requires at least a few tens of thousand of voids (of similar size and shape). Given the fact that for larger redshift (for tens of thousands of voids, one requires higher redshift to probe a large volume) the amplitude of $\Delta$ is lower this shows that a direct method of detecting this feature based on the number counts of galaxies within voids is not effective. For example, the void catalog of SDSS DR12 BOSS\footnote{The Sloan Digital Sky Survey, \url{https://www.sdss.org/}} consists of 10,643 voids \cite{2017ApJ...835..161M}, discovered between $z =  0.2$ and $0.7$. While impressive, it still seems insufficient to detect the discussed feature (indeed \cf Fig. 6 in  \cite{2017ApJ...835..161M}). 
A potentially better strategy for detecting this feature is based on the indirect method of gravitational lensing \cite{2013ApJ...762L..20K,2021arXiv211014089A,2021tbp} which could also be combined with the Doppler lensing method  \cite{2013PhRvL.110b1302B,2014MNRAS.443.1900B,2021tbp}.
The detectability issue will be explored in future work.

\section{Conclusion}
\label{sec:conclusion}

The aim of this paper has been to investigate the effects of DDM on cosmic voids. 
We focused on cosmic voids as they are pristine environments that are less prone to contamination by complex baryonic physics than to dense environments such as galaxy clusters. By targeting voids one hopes to disentangle the observational signatures of dark matter from those of baryonic matter. 
To this end, we have developed a novel method of modelling the decay of dark matter. In this model the comoving CDM component decays into another dark particle with a non-relativistic velocity, thereby producing a secondary fluid which has non-comoving average spatial velocity induced by the velocity of the produced daughter particles. The effect of this non-comoving secondary fluid on the evolution of cosmological voids has been investigated. 

Recent work, \cite{gaspar2019non, tsagas2010large}, has presented thorough analyses of the effect of non-comoving components upon cosmological and structure formation. The present article has attempted to further these works by suggesting unstable DDM as a mechanism for producing non-comoving fluid components. Rather than the non-comoving velocity distribution, we have explored the effect of the DDM particle parameters on void evolution. The results of this analysis suggest that for S-type voids there is the formation of a significant secondary substructure induced by the non-comoving DM component produced via the decay.

With further sophistication, the model proposed and presented here may offer a contribution towards constraining the unknown particle properties of DM, in conjunction with constraints determined by other means. Our method could be used to further constrain or discover DDM signatures. For example, if no secondary substructure is observed than one could rule out combination of decay lifetimes $\Gamma < H_{0}^{-1}$ Gyr and injection velocities of $v_{i} > 10$ km/s. Our methods will allow us to put tighter constraints on DDM.
As a preliminary investigation, our goal has been to proffer cosmological voids as a possible means of indirect detection of DDM, with no attempts made to constrain DDM particle candidates, nor to satisfy the possible thermodynamic relations which would be required for physicality, \cite{coley1989thermodynamics}. The present paper gives a proof-of-concept that demonstrates that cosmological voids can offer an indirect, observational means to constrain properties of DM.

\section*{Acknowledgements}

The authors wish to acknowledge and thank 
Andrew Bassom, Larry Forbes, and Stephen Walters 
for their support and discussions. 
KB acknowledges support from the Australian Research Council through the Future Fellowship FT140101270.

\bibliographystyle{apsrev}

\end{document}